\documentclass[%
superscriptaddress,
amsmath,amssymb,
twocolumn,
]{revtex4}
\usepackage[
colorlinks=true,
linkcolor=blue,
filecolor=blue,
citecolor=blue,
urlcolor=blue,
]{hyperref}
\usepackage{graphicx}
\usepackage{siunitx}
\usepackage{soul}
\setstcolor{red}
\usepackage[normalem]{ulem}

\newcommand{\add}[1]{#1}
\newcommand{\erase}[1]{}
\newcommand{\newadd}[1]{#1}
\newcommand{\newerase}[1]{}
\newcommand{\neweraseforref}[1]{}

\begin{document}

\title{
    \add{\newerase{Multi-Step}\newadd{Sequential and} Programmable Squeezing Gates}\\\add{for Optical Non-Gaussian Input States}
}

\author{Takato Yoshida}
\affiliation{
    Department of Applied Physics, School of Engineering, The University of Tokyo,
    7-3-1 Hongo, Bunkyo-ku, Tokyo 113-8656, Japan
}
\author{Daichi Okuno}
\affiliation{
    Department of Applied Physics, School of Engineering, The University of Tokyo,
    7-3-1 Hongo, Bunkyo-ku, Tokyo 113-8656, Japan
}
\author{Takahiro Kashiwazaki}
\affiliation{
    NTT Device Technology Labs, NTT Corporation,
    3-1, Morinosato Wakamiya, Atsugi, Kanagawa 243-0198, Japan
}
\author{Takeshi Umeki}
\affiliation{
    NTT Device Technology Labs, NTT Corporation,
    3-1, Morinosato Wakamiya, Atsugi, Kanagawa 243-0198, Japan
}
\author{Shigehito Miki}
\affiliation{
    Advanced ICT Research Institute, National Institute of Information and Communications Technology, 
    588-2 Iwaoka, Nishi\add{-ku}, Kobe 651-2492, Japan
}
\author{Fumihiro China}
\affiliation{
    Advanced ICT Research Institute, National Institute of Information and Communications Technology, 
    588-2 Iwaoka, Nishi\add{-ku}, Kobe 651-2492, Japan
}
\author{Masahiro Yabuno}
\affiliation{
    Advanced ICT Research Institute, National Institute of Information and Communications Technology, 
    588-2 Iwaoka, Nishi\add{-ku}, Kobe 651-2492, Japan
}
\author{Hirotaka Terai}
\affiliation{
    Advanced ICT Research Institute, National Institute of Information and Communications Technology, 
    588-2 Iwaoka, Nishi\add{-ku}, Kobe 651-2492, Japan
}
\author{Shuntaro Takeda}
\email{takeda@ap.t.u-tokyo.ac.jp}
\affiliation{
    Department of Applied Physics, School of Engineering, The University of Tokyo,
    7-3-1 Hongo, Bunkyo-ku, Tokyo 113-8656, Japan
}
\begin{abstract}
    \noindent Quantum computing has been pursued with various hardware platforms, and an optical system is one of the most reasonable choices for large-scale computation. In the optical continuous-variable computation scheme, the incorporation of Gaussian gates and a highly non-classical non-Gaussian state enables universal quantum computation. Although basic technologies for Gaussian gates and non-Gaussian state generations have long been developed, these building blocks have not yet been integrated in a scalable fashion. Here, we integrate them to develop a scalable and programmable optical quantum computing platform that can sequentially perform an essential Gaussian gate, the squeezing gate, on a non-Gaussian input state. The key enablers are a loop-based optical circuit with dynamical and programmable controllability and its time-synchronization with the probabilistic non-Gaussian state generation. We verify the deterministic, programmable, and repeatable quantum gates on a typical non-Gaussian state by implementing up to three-step gates. The gates implemented are so high-quality that strong evidence of the states' non-classicalities, negativities of the Wigner functions, are preserved even after multistep gates. This platform is compatible with other non-Gaussian states and \erase{realizes}\add{can in principle realize} large-scale universal quantum computing by incorporating other existing processing technologies.
\end{abstract}
\maketitle

\section{Introduction}
\noindent While various physical systems have emerged for quantum computing~\cite{bruzewiczTrappedion,googlequantumaiSuppressing,winterspergerNeutral}, an optical system employing the continuous-variable (CV) scheme has been attracting much interest for its strength in scalability~\cite{takedaLargescale}. The optical CV scheme, which utilizes the quadrature amplitude of the light field for quantum computing, has offered deterministic quantum gates~\cite{yoshikawaDemonstration,ukaiDemonstration,suGate} with its practical advantage lying in the unconditionally prepared ancillary squeezed vacuum and the highly efficient homodyne measurement. A promising idea to integrate the deterministic gates to build a scalable quantum computing platform is to adopt the time-domain multiplexing approach unique to an optical system. In this approach, the quantum information is encoded in sequential optical pulses on a single (or a few) path(s), and the same optical components are repeatedly used to process them at different times. This approach, in fact, has recently led to the development of several platforms able to perform programmable multi-step quantum gates on a single- or multi-mode input state~\cite{asavanantTimeDomainMultiplexed,larsenDeterministic,enomotoProgrammable,yonezuTimeDomain}.

However, these scalable platforms are currently limited to performing calculations within the Gaussian realm, which can be efficiently simulated by a classical computer~\cite{bartlettEfficient}. This limitation implies a lack of the ability to perform universal quantum computation. The missing piece for achieving universality is a non-Gaussian quantum state. When a specific non-Gaussian ancillary state \add{known as a cubic phase state} is injected, Gaussian platforms, which consist of optical Gaussian gates and homodyne measurements, enable universal computation including a non-Gaussian gate~\cite{miyataImplementation}\erase{, even suffice for fault tolerance}\add{. Furthermore, fault tolerance is also attainable on these platforms by injecting non-Gaussian qubit states, specifically Gottesman-Kitaev-Preskill (GKP) qubits}~\cite{baragiolaAllGaussian,yamasakiCostreduced}. Nevertheless, the realization of scalable Gaussian platforms compatible with non-Gaussian states had long been obstructed by technical challenges arising from the probabilistic generations of such states: non-Gaussian states thus far had only undergone single-step gates without electrical programmability in a non-scalable fashion~\cite{miwaExploring,wangExperimental}. Only recently, time-synchronization of a dynamical optical circuit with a probabilistic non-Gaussian source has been reported as a piece for a scalable platform~\cite{okunoTimedomain}.

In this work, as a crucial step toward simultaneously achieving scalability and universality, we develop an optical quantum computing platform that can repeatedly and programmably perform a fundamental Gaussian gate, the squeezing gate, on a non-Gaussian input state. This platform in principle provides arbitrarily many steps of squeezing gates with a single setup, and the gates are electrically programmable step by step. This is achieved by incorporating two major components. One is a scalable and programmable loop-based processor~\cite{enomotoProgrammable}. The other is a time-domain-multiplexed light source unit that prepares a train of optical pulses consisting of the non-Gaussian input state \add{(Schr\"{o}dinger\newerase{'s} cat state)} and the ancillary states for the gate, which is sent to the processor. Time-synchronizing these two components with the probabilistic non-Gaussian state generation enables performing the gate known as the measurement-induced squeezing~\cite{filipMeasurementinduced} on that state. \add{This is an essential gate that composes a universal gate set and also serves as the foundation for various other measurement-induced gates.} We verify the deterministic, programmable, and repeatable squeezing gates by implementing various single- or multi-step gates. In addition, the high qualities of the gates are supported by the remaining negativities of the Wigner functions, which are widely recognized criteria for the non-classicalities of the quantum states, even after two-step gates without any loss correction. The squeezing gate can be combined with other easier-to-implement gates to enable any single-mode or multi-mode Gaussian gate, suggesting a versatile testbed for quantum computing with a non-Gaussian input as a near-term application of our platform. Moreover, since a non-Gaussian gate, which completes the universal gate set, can be realized through a similar measurement-based gate protocol with an alternative non-Gaussian ancilla~\cite{miyataImplementation,enomotoProgrammable}, our work will lead to fully universal quantum computing. \add{\newerase{In the context of fault tolerance, as the Schr\"{o}dinger's cat states can ultimately be transformed into GKP qubits using Gaussian operations~[20--22], our system is a promising platform that contains all the minimal elements needed for fault-tolerant universal computing.}}

\begin{figure*}[htbp]
    \centering
    \includegraphics[scale=1]{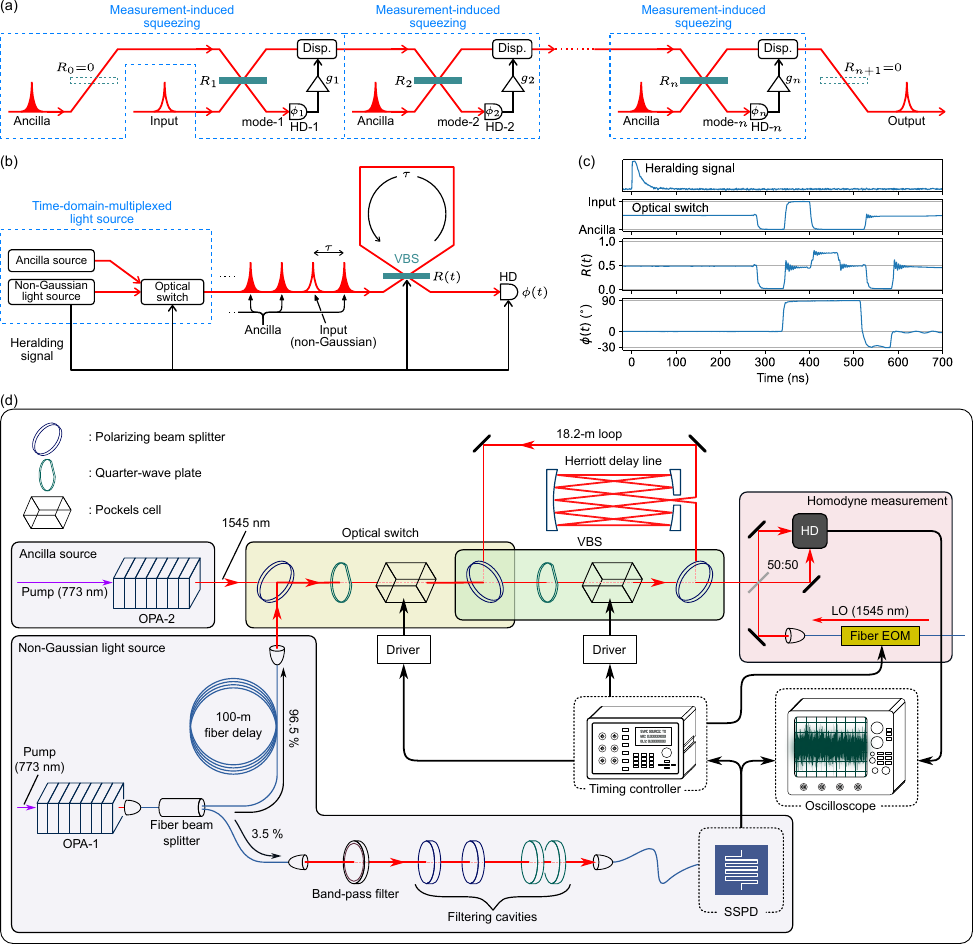}
    \caption{Conceptual diagram of our platform and experimental setup. (a) Typical implementation of a multi-step measurement-induced squeezing based on path encoding. Disp., displacement operation; HD, homodyne detector. (b) Conceptual diagram of our platform. VBS, variable beam splitter. (c) Timing chart of each component for the third gate in Table \ref{tab: condition2}. Zero on the horizontal axis corresponds to the rising of the heralding signal. (d) Experimental setup. EOM, electro-optic modulator; LO, local oscillator; OPA, optical parametric amplifier; SSPD, superconducting nanostrip single-photon detector.}
    \label{fig: concept}
\end{figure*}

\section{Results}
\subsection{Our platform for a multi-step squeezing gate}
\noindent Before describing our platform, we begin by briefly introducing the measurement-induced squeezing gate~\cite{filipMeasurementinduced} with Fig. \ref{fig: concept}(a), which illustrates a typical schematic of a multi-step gate. The single-step $\hat{x}$-squeezing is composed of the following steps, where $\hat{x}$ and $\hat{p}$ represent the orthogonal quadrature operators of the light field: (1) an arbitrary input state is coupled to an ancillary $\hat{x}$-squeezed vacuum at a beam splitter with reflectivity $R$ and transmissivity $T$ ($R+T=1$); (2) the $\hat{p}$ quadrature of one of the resultant modes is measured by the homodyne detector (HD); (3) the measurement outcome $m$ is fed forward to displace the $\hat{p}$ quadrature of the other mode by $gm$ to cancel out the antisqueezing component of the ancillary state, where $g=\sqrt{T/R}$ is called the feedforward gain. In the ideal limit of infinitely squeezed ancilla, this gate transforms the quadrature operators of the input state as $\hat{x}\mapsto\sqrt{R}\hat{x}$ and $\hat{p}\mapsto\hat{p}/\sqrt{R}$ in the Heisenberg picture. Note that this input-output relation changes to $\hat{x}\mapsto\sqrt{T}\hat{x}$ and $\hat{p}\mapsto\hat{p}/\sqrt{T}$ with $g=-\sqrt{R/T}$ only for the first step in Fig. \ref{fig: concept}(a) due to the inverse positioning of the input and the ancilla. These transformations correspond to the squeezing gates, and the degree of squeezing depends only on the beam-splitter reflectivity $R$. It is worth noting that the quadrature being squeezed, $\hat{x}$ in the above example, can be controlled by adjusting the phase of the ancilla with respect to the input and accordingly changing the two quadratures being measured or displaced.

Figure \ref{fig: concept}(b) shows a conceptual diagram of our platform, which is markedly more scalable while having exactly the same functionality as the circuit in Fig. \ref{fig: concept}(a) has. In this architecture, the optical pulses are arranged in the time domain, and a single dynamical processing unit is repeatedly used to sequentially process them in a resource-efficient manner. This platform is composed of three parts: the time-domain-multiplexed light source, the loop-based optical circuit, and the HD. The individual properties or functionalities of them are as follows. The light source plays the role of preparing the optical pulses arranged in the time domain: the constituent optical switch sequentially transfers either a probabilistically generated non-Gaussian input state or a deterministically generated ancillary state to the loop-based circuit with a time separation of $\tau$. The loop-based circuit has an optical delay line with a round-trip time of $\tau$, which makes the incoming two adjacent optical pulses coincide at a variable beam splitter (VBS) for interference. At the downstream of the loop part, the HD with a variable measurement basis $\phi(t)$ is placed to measure $\hat{x}\cos\phi(t)+\hat{p}\sin\phi(t)$ of the incident beam. All the dynamical parameters---switching status, VBS reflectivity $R(t)$, homodyne measurement basis $\phi(t)$---are electrically programmable. Notably, as a platform for a non-Gaussian input, these parameters are integrally controlled in synchronization with an electric signal that heralds the generation of the non-Gaussian state. Figure \ref{fig: concept}(c) shows a typical timing chart.

This platform performs an $n$-step squeezing gate in the following way. When a non-Gaussian input state is generated, the optical switch sends it to the loop part with ancillae neighboring, forming a pulse train in the order in Fig. \ref{fig: concept}(b). In the loop part, the reflectivity of the VBS is firstly set to zero ($R_0=0$ in Fig. \ref{fig: concept}(a), which depicts an equivalent circuit of our platform) for the loop to take in the ancilla. To avoid the fragile non-Gaussian state suffering excess loss in the loop, the ancilla should be thrown into the loop ahead for the first step out of the $n$ steps. The VBS reflectivity then sequentially takes $n$ values ($R_1,\ \dots,\ R_n$) to act as $n$ different beam splitters. At the same time, the HD continuously measures one of the VBS outputs with its basis $\phi(t)$ dynamically controlled for the desired gate. This continuous measurement can be split into multiple measurements ($\mathrm{HD-}1,\ \dots,\ \mathrm{HD-}n$) on different time bins. Finally, the VBS is again made transparent ($R_{n+1}=0$) to let the output state exit the loop, and the HD measures it for state characterization. The full outcome of the homodyne measurement is stored in a classical computer, and all the displacement operations are performed by the numerical post-processing on it. Owing to the dynamical controllability and the loop-based structure enabling sequential processing, our platform can perform arbitrarily many steps of squeezing gates with a constant number of hardware components, offering scalability.

Figure \ref{fig: concept}(d) depicts the physical setup. Any non-Gaussian input state is acceptable as long as the heralding signal arises, and we choose the Schr\"{o}dinger\newerase{'s} cat state for the demonstration. For the state preparation with the photon subtraction scheme~\cite{neergaard-nielsenGeneration,wakuiPhoton}, a squeezed vacuum is produced by an optical parametric amplifier (OPA) and subsequently thrown into a fiber beam splitter (FBS) one output of which leads to a superconducting nanostrip single-photon detector (SSPD). Photon detection on the SSPD generates the Schr\"{o}dinger\newerase{'s} cat on the other output of the FBS in a specific wave packet defined by filtering optics before the SSPD. The ancillary squeezed vacuum is continuously produced from another OPA. Except for these light sources, the whole optical system is constructed in free space including the loop-based circuit most of which is folded in the mechanically stable Herriott-type delay line~\cite{herriottOffAxis}. The output state of the gate is fully characterized by the homodyne tomography~\cite{lvovskyIterative} with neither post-selection nor loss correction. The timing controller synchronizes the entire system with the photon detection signal from the SSPD at a nanosecond time scale. The detail of the setup is described in Appendix \ref{app: experimental setup}.

\begin{table*}[htbp]
    \caption{Working conditions and fidelities for the single-step squeezing gates. ``$\hat{x}$-sq.'' (``$\hat{p}$-sq.'') in the ``Ancilla'' column refers to the $\hat{x}$-squeezed vacuum ($\hat{p}$-squeezed vacuum). $R_1$, $\phi_1$, and $g_1$ represent the VBS reflectivity, the measurement angle, and the feedforward gain, respectively. These symbols can be found in Fig. \ref{fig: concept}(a), which depicts an equivalent circuit of our platform. The definitions of the fidelities, $F_{\mathrm{real}}$\erase{ and}\add{,} $F_{\mathrm{ideal}}$\add{, and $F_{\mathrm{ideal}}^{\mathrm{(theory)}}$}, are given in the text. The center values and the error ranges are estimated as averages and standard errors when splitting the whole \SI{36000} data into five independent subsets, respectively. The same is the case for Table \ref{tab: condition2}.}
    \label{tab: condition1}
    \begin{center}
      \begin{tabular}{lc|cccc|cc|c} \hline
        \multicolumn{2}{c|}{Gate} & \multicolumn{4}{c|}{Working condition} & \multicolumn{2}{c|}{\begin{tabular}{c}\erase{Result}\add{Experimental}\\\add{fidelities}\end{tabular}} & \begin{tabular}{c}\add{Theoretical}\\\add{fidelity}\end{tabular} \\ \hline
        $r$ & \begin{tabular}{c}Squeezed\\quadrature\end{tabular} & Ancilla & $R_1$ & $\phi_1$ & $g_1$ & $F_{\mathrm{real}}$ & $F_{\mathrm{ideal}}$ & \add{$F_{\mathrm{ideal}}^{\mathrm{(theory)}}$} \\ \hline\hline
        \num{0.26} & $\hat{x}$ & $\hat{x}$-sq. & \num{0.40} & \ang{90} & \num{-0.82} & \num{0.976(3)} & \num{0.956(2)} & \add{\num{0.961}} \\
        \num{0.46} & $\hat{x}$ & $\hat{x}$-sq. & \num{0.60} & \ang{90} & \num{-1.22} & \num{0.956(4)} & \num{0.907(3)} & \add{\num{0.907}} \\
        \num{-0.26} & $\hat{p}$ & $\hat{p}$-sq. & \num{0.40} & \ang{0} & \num{-0.82} & \num{0.9783(9)} & \num{0.959(3)} & \add{\num{0.961}} \\
        \num{-0.46} & $\hat{p}$ & $\hat{p}$-sq. & \num{0.60} & \ang{0} & \num{-1.22} & \num{0.979(3)} & \num{0.904(4)} & \add{\num{0.911}} \\ \hline
      \end{tabular}
    \end{center}
\end{table*}

\begin{figure*}[htbp]
    \centering
    \includegraphics[scale=1]{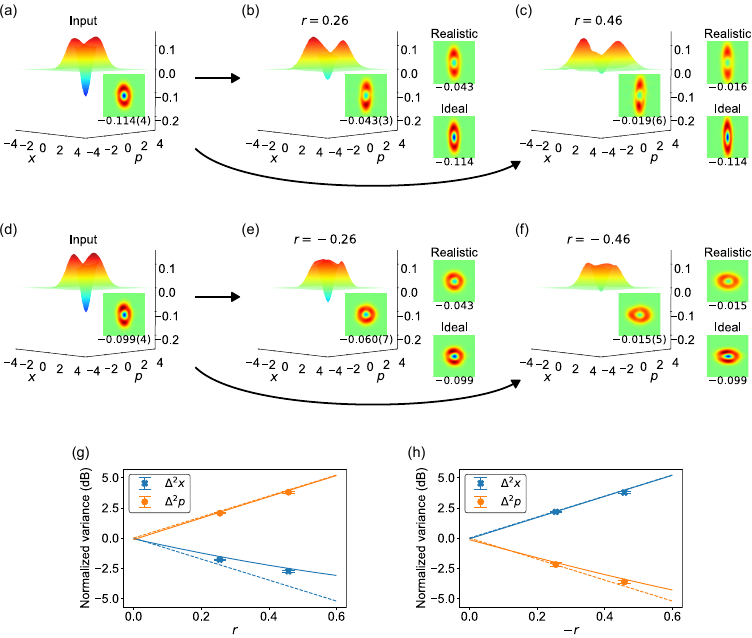}
    \caption{State characterizations for the single-step squeezing gates. (a),(d) Wigner function of the input Schr\"{o}dinger\newerase{'s} cat state for the squeezing with positive and negative squeezing parameters, respectively (they are identical in principle). The convention $\hbar=1$ is adopted. (b),(c),(e),(f) Wigner function of the experimentally obtained output state and the theoretically derived ones. The horizontal and the vertical directions of the inset square panel correspond to $x$ and $p$, respectively. The figure below each panel represents the value of $W(0,0)$. See the text for the description of the theoretical predictions. (g),(h) Normalized variances as a function of a target squeezing parameter. The solid and the dashed lines are derived from the realistic and the ideal models, respectively. See the text for the definitions of the normalized variances. For all items, the presented center values and the error ranges are estimated as averages and standard errors when splitting the whole \SI{36000} data into five independent subsets, respectively. The same is the case for Fig. \ref{fig: result2}.}
    \label{fig: result1}
\end{figure*}

\subsection{Single-step gate} \label{subsec: single-step gate}
\noindent For the certification of the programmable and deterministic gates of our platform, we implement single-step squeezing gates with various squeezing parameters. A squeezing gate is characterized by an operator $\hat{S}(r)$ that transforms quadrature operators as $\hat{S}^{\dagger}\hat{x}\hat{S}(r)=e^{-r}\hat{x}$ and $\hat{S}^{\dagger}\hat{p}\hat{S}(r)=e^{r}\hat{p}$. We call $r$ a squeezing parameter, and we choose $r=\pm0.26$, $\pm0.46$ for the demonstration. The system working condition for each gate is listed in Table \ref{tab: condition1}.

Figure \ref{fig: result1} and Table \ref{tab: condition1} summarize the experimental results. We first measured the time-domain-multiplexed input Schr\"{o}dinger\newerase{'s} cat and the ancillary squeezed vacuum by setting the VBS maximally reflecting, before implementing the squeezing gates. The reconstructed Wigner functions of the input cats for the squeezing gates with positive and negative squeezing parameters are shown in Figs. \ref{fig: result1}(a) and \ref{fig: result1}(d), respectively, which are identical in principle. The squeezing levels of the ancillae were measured to be \SI{\sim-4}{\decibel} (see Appendix \ref{app: theoretical models} for more precise figures). We then present the output Wigner functions of the squeezing gates in Figs. \ref{fig: result1}(b), \ref{fig: result1}(c), \ref{fig: result1}(e), and \ref{fig: result1}(f). Alongside each experimental outcome, the theoretical ones derived from the realistic or ideal model are shown with the corresponding label attached. The realistic model involves the finitely squeezed ancilla and the optical loss whereas the ideal one has the infinitely squeezed ancilla and no optical loss (the detail is described in Appendix \ref{app: theoretical models}). All the experimental outcomes graphically agree well with the realistic predictions.

For quantitative evaluation, we employ three different figures of merit: quadrature variances representing the degree of squeezing, Wigner negativity representing the quality of the gate, and fidelities comprehensively representing the circuit performance. We first present the quadrature variances of the output states in Figs. \ref{fig: result1}(g) and \ref{fig: result1}(h). They are normalized by those of the corresponding input state: in the specific form, $\Delta^2q$ $(q=x,p)$ is defined as $(\langle\hat{q}_{\mathrm{out}}^2\rangle-\langle\hat{q}_{\mathrm{out}}\rangle^2)/(\langle\hat{q}_{\mathrm{in}}^2\rangle-\langle\hat{q}_{\mathrm{in}}\rangle^2)$, where $\langle\cdots\rangle$ denotes the mean value and the subscripts ``out'' and ``in'' refer to the output and the input states, respectively. For four different target squeezing parameters, the experimentally obtained variances have little deviations from the realistic predictions, indicating that our platform properly performs squeezing gates in a programmable manner. Next, we show the Wigner negativities of the output states, which is defined as the value of $W(0,0)$ for the output Wigner function $W(x,p)$, below the square panels in Fig. \ref{fig: result1}. This value, in our case, represents the quality of the gate. This is because the negative $W(0,0)$ in general signifies the non-classicality of the state and is degraded through the gate process whereas the ideal squeezing keeps it constant. All the experimental results show clear negative values, revealing the high qualities of the gates of our platform. Finally, as comprehensive indicators, we present two fidelities $F_{\mathrm{real}}$ and $F_{\mathrm{ideal}}$ in Table \ref{tab: condition1}. $F_{\mathrm{real}}$ ($F_{\mathrm{ideal}}$) is the fidelity~\cite{jozsaFidelity} between the experimental output and the theoretical one derived from the realistic (ideal) model described above. \add{We also include the theoretical fidelity $F_{\mathrm{ideal}}^{\mathrm{(theory)}}$ in the same table for reference. This value, obtained numerically, represents the fidelity between the theoretical outputs derived from the realistic and the ideal models, predicting $F_{\mathrm{ideal}}$.} For all gates implemented, both $F_{\mathrm{ideal}}$ and $F_{\mathrm{real}}$ are reasonably high. Non-unity values of $F_{\mathrm{real}}$ can be attributed to deviations between the simple theoretical model and the actual experimental situation, such as the spatial mode mismatch at the VBS between the incoming and the circulated beams.

As a whole, Fig. \ref{fig: result1} and Table \ref{tab: condition1} graphically and quantitatively demonstrate that our platform deterministically performs high-quality squeezing gates on the non-Gaussian cat state in a programmable manner. We have also evaluated each gate with a different input state, the vacuum field, and confirmed that the experimental output is again consistent with the theoretical prediction. The result is shown in Table S1 and Fig. S1 in Supplemental Material~\cite{ourSupplementalMaterial}.

\subsection{Multi-step gate}
\noindent For the demonstration of our platform's ability to perform multi-step gates, we implement up to three-step squeezings. The squeezing parameters are chosen to be $(r_1,r_2,r_3)=\pm(\num{0.33},\num{0.14},\num{0.37})$, where $r_i$ refers to the one for the $i$th step. The working conditions are summarized in Table \ref{tab: condition2}.

\begin{figure*}[htbp]
    \centering
    \includegraphics[scale=1]{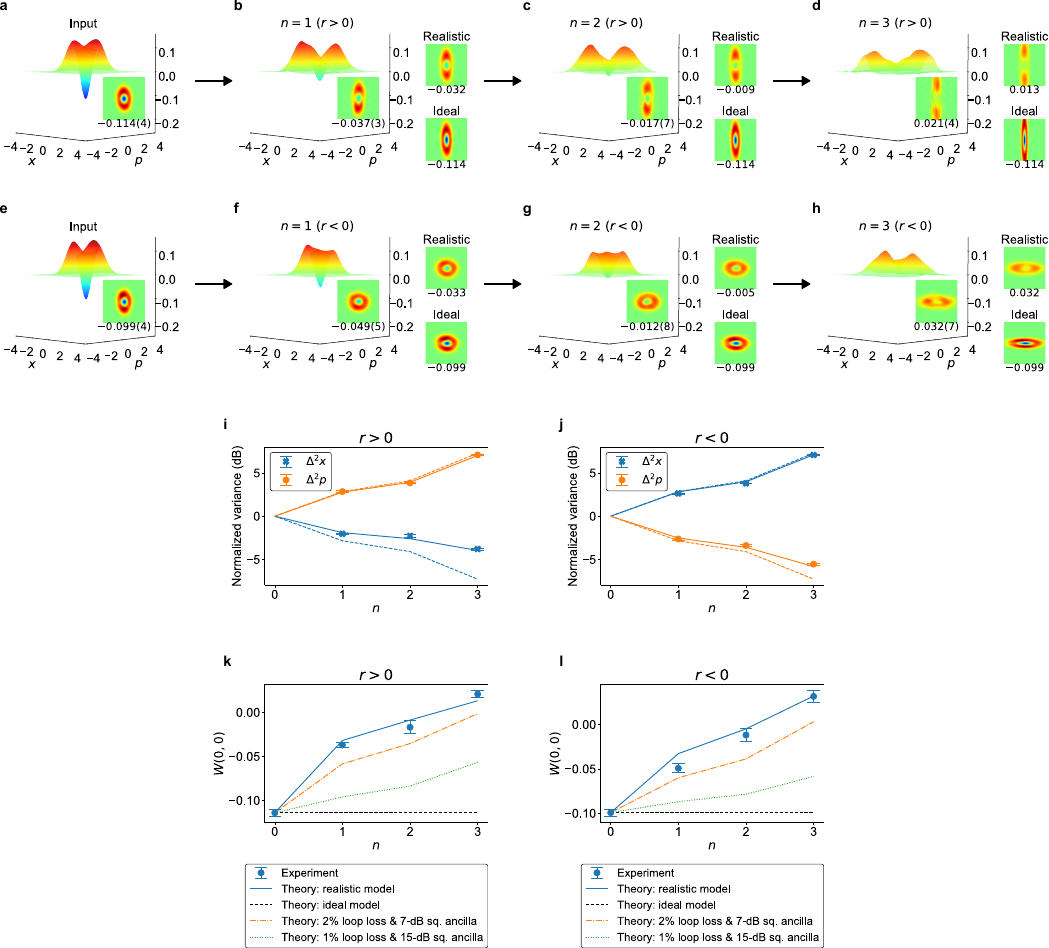}
    \caption{State characterizations for the multi-step squeezing gates. (a),(e) Wigner function of the input Schr\"{o}dinger\newerase{'s} cat state for the squeezing with positive and negative squeezing parameters, respectively. These are the same figures as in Fig. \ref{fig: result1}. The convention $\hbar=1$ is adopted. (b),(c),(d),(f),(g),(h) Wigner function of the experimentally obtained output state and the theoretically derived ones. $n$ denotes the number of steps. (i),(j) Normalized variances as a function of the number of steps. \newadd{(k),(l) $W(0,0)$ values of the output $W(x,p)$ as a function of the number of steps. The solid blue and dashed black lines are drawn using the realistic and ideal models described in the text, respectively. The dashdot orange line (dotted green line) is obtained with a model with \SI{2}{\percent} (\SI{1}{\percent}) loop loss and an ancilla of \SI[number-unit-product=\text{-}]{7}{\decibel} (\SI[number-unit-product=\text{-}]{15}{\decibel}) squeezed vacuum, where ``sq.'' in the legend denotes ``squeezed.''} \newerase{All items}\newadd{Figures (a)--(j)} are shown in the same manner as in Fig. \ref{fig: result1} except that (i) and (j) are shown as functions of $n$.}
    \label{fig: result2}
\end{figure*}

Figure \ref{fig: result2} and Table \ref{tab: condition2} present the step-by-step results similarly to Fig. \ref{fig: result1} and Table \ref{tab: condition1}. The output Wigner functions in Figs. \ref{fig: result2}(b)--\ref{fig: result2}(d) and \ref{fig: result2}(f)--\ref{fig: result2}(h), along with the quadrature variances in Figs. \ref{fig: result2}(i) and \ref{fig: result2}(j), show that the experimental results graphically and quantitatively agree well with the theoretical predictions. Furthermore, the two fidelities, including $F_{\mathrm{ideal}}$, in Table \ref{tab: condition2} remain at a relatively high level over the multiple steps. These results collectively demonstrate that our platform reliably performs multi-step squeezings while programmably changing the squeezing parameter. The preserved Wigner negativities for up to two steps highlight the qualities of the gates. \newadd{Figures \ref{fig: result2}(k) and \ref{fig: result2}(l) show the evolution of negativity as a function of the number of gates, along with theoretical lines derived from the realistic and ideal models, as well as models under improving system parameters. A detailed description of these additional models can be found in Appendix \ref{app: optical loss}, in the context of mitigating optical losses.} \newerase{The primary cause of the negativity's degradation, as well as $F_{\mathrm{ideal}}$'s degradation, is the finitely squeezed quantum noise of the ancilla rather than the optical loss in the loop part, which is as low as 4 \%. This suggests that higher-quality gates are attainable in this architecture simply by further suppressing the ancilla's squeezed quantum noise. }\add{\newerase{A more detailed discussion of optical losses can be found in Appendix }\neweraseforref{\ref{app: optical loss}.} }It should be emphasized here that the available number of steps in \newerase{this}\newadd{our} architecture is not restricted to three but infinite in principle.

Similarly to the case of single-step gates, we have applied the same gates to the vacuum field and presented the result in Table S2 and Fig. S2 in Supplemental Material~\cite{ourSupplementalMaterial}, which again proves the validity of the gates.

\begin{table*}[htbp]
    \caption{Working conditions and fidelities for the multi-step squeezing gates. All items are listed in the same manner as in Table \ref{tab: condition1}. The symbols $R_i$, $\phi_i$, and $g_i$ $(i=1,2,3)$ can be found in Fig. \ref{fig: concept}(a).}
    \label{tab: condition2}
    \begin{center}
      \begin{tabular}{lc|clll|cc|c} \hline
        \multicolumn{2}{c|}{Gate} & \multicolumn{4}{c|}{Working condition} & \multicolumn{2}{c|}{\begin{tabular}{c}\erase{Result}\add{Experimental}\\\add{fidelities}\end{tabular}} & \begin{tabular}{c}\add{Theoretical}\\\add{fidelity}\end{tabular} \\ \hline
        $(r_1,r_2,\dots)$ & \begin{tabular}{c}Squeezed\\quadrature\end{tabular} & Ancillae & $(R_1,R_2,\dots)$ & $(\phi_1,\phi_2,\dots)$ & $(g_1,g_2,\dots)$ & $F_{\mathrm{real}}$ & $F_{\mathrm{ideal}}$ & \add{$F_{\mathrm{ideal}}^{\mathrm{(theory)}}$} \\ \hline\hline
        $(\num{0.33})$ & $\hat{x}$ & $\hat{x}$-sq. & $(\num{0.48})$ & $(\ang{90})$ & $(\num{-0.96})$ & \num{0.968(2)} & \num{0.939(4)} & \add{\num{0.944}}\\
        $(\num{0.33},\num{0.14})$ & $\hat{x}$ & $\hat{x}$-sq. & $(\num{0.48},\num{0.75})$ & $(\ang{90},\ang{90})$ & $(\num{-0.96},\num{0.58})$ & \num{0.958(4)} & \num{0.876(4)} & \add{\num{0.905}}\\
        $(\num{0.33},\num{0.14},\num{0.37})$ & $\hat{x}$ & $\hat{x}$-sq. & $(\num{0.48},\num{0.75},\num{0.48})$ & $(\ang{90},\ang{90},\ang{90})$ & $(\num{-0.96},\num{0.58},\num{1.04})$ & \num{0.930(3)} & \num{0.753(6)} & \add{\num{0.793}}\\
        $(\num{-0.33})$ & $\hat{p}$ & $\hat{p}$-sq. & $(\num{0.48})$ & $(\ang{0})$ & $(\num{-0.96})$ & \num{0.981(2)} & \num{0.948(5)} & \add{\num{0.945}}\\
        $(\num{-0.33},\num{-0.14})$ & $\hat{p}$ & $\hat{p}$-sq. & $(\num{0.48},\num{0.75})$ & $(\ang{0},\ang{0})$ & $(\num{-0.96},\num{0.58})$ & \num{0.978(3)} & \num{0.888(5)} & \add{\num{0.910}}\\
        $(\num{-0.33},\num{-0.14},\num{-0.37})$ & $\hat{p}$ & $\hat{p}$-sq. & $(\num{0.48},\num{0.75},\num{0.48})$ & $(\ang{0},\ang{0},\ang{0})$ & $(\num{-0.96},\num{0.58},\num{1.04})$ & \num{0.971(2)} & \num{0.770(7)} & \add{\num{0.821}}\\ \hline
      \end{tabular}
    \end{center}
\end{table*}

\section{Discussion}
\noindent In conclusion, we have developed the optical quantum computing platform that can repeatedly and programmably perform a measurement-induced squeezing gate on a non-Gaussian state. We have verified the deterministic, programmable, and repeatable nature of the gates. Our platform can deal with not only the Schr\"{o}dinger\newerase{'s} cat state but also any other non-Gaussian states as long as the heralding signal arises, including the cubic phase state~\cite{yukawaEmulating} and the \erase{Gottesman-Kitaev-Preskill}\add{GKP} state~\cite{konnoLogical}. In addition to the squeezing gate on input non-Gaussian states, the current platform has the potential to acquire another functionality when a specific non-Gaussian state, instead of the cat, is injected as an ancilla. Notably, the ancillary cubic phase state, together with the already reported nonlinear feedforward system~\cite{sakaguchiNonlinear}, enables the implementation of the cubic phase gate~\cite{enomotoProgrammable}. Further adding a phase shifter inside the loop evolves the platform to a single-mode universal one since an arbitrary single-mode gate is composed of displacement, squeezing, phase shift, and the cubic phase gate~\cite{lloydQuantum}. The main technical challenge, deterministic processing of the non-Gaussian state in the time domain, has been addressed in this work; the remaining task is the integration of these technologies. Furthermore, our platform can be extended for multi-mode processing~\cite{takedaUniversal} and eventually leads to fault-tolerant computing, which only necessitates Gaussian platforms once the proper non-Gaussian qubit states are supplied~\cite{baragiolaAllGaussian,yamasakiCostreduced}. Therefore, our work represents a fundamental milestone toward the realization of scalable, universal, and fault-tolerant quantum computing.

The data that support the findings of this study are available from the corresponding author upon reasonable request.

\section*{ACKNOWLEDGEMENTS}
\noindent This work was partly supported by
JST Grant Numbers JPMJFR223R, JPMJMS2064, and JPMJPF2221,
JSPS KAKENHI Grant Numbers 23H01102 and 23K17300,
the Canon Foundation,
and MEXT Leading Initiative for Excellent Young Researchers.

T.Y. and S.T. designed the whole system. T.Y. and D.O. contributed to the development and debugging of the experimental setup, the data acquisition and analysis, and the numerical simulation. T.K. and T.U. supplied the OPA modules. S.M., H.T., M.Y., and F.C. supplied the SSPD. S.T. conceived, planned, and supervised the project. T.Y. wrote the manuscript with assistance from D.O., S.T., and other co-authors.

\appendix
\section{Experimental setup} \label{app: experimental setup}
\noindent Figure \ref{fig: concept}(d) illustrates the experimental setup, which is an extended one from that in our previous work~\cite{okunoTimedomain}. Refer to Ref.~\cite{okunoTimedomain} for more information on the common part.

We use a continuous-wave laser with a wavelength of \SI{1545}{\nano\metre}. Frequency-doubled beams pump two OPAs~\cite{kashiwazakiFabrication,kashiwazakiOver8dB} to produce squeezed vacua, from one of which a photon is subtracted to obtain the non-Gaussian Schr\"{o}dinger\newerase{'s} cat state. A band-pass filter and two filtering cavities before the SSPD define the temporal wave packet $f(t)$ of the cat state as
\begin{equation}
    f(t) \propto \left(e^{\gamma_1(t-t_0)}-e^{\gamma_2(t-t_0)}\right)\Theta(t_0-t), \label{eq: modefunction}
\end{equation}
where $\gamma_1$ and $\gamma_2$ are cavities' bandwidth (half width at half maximum) and $t_0$ is the photon detection timing on the SSPD. $\Theta$ denotes the Heaviside step function. The configuration of the VBS is the same as those in our previous works~\cite{takedaOndemand,enomotoProgrammable,yonezuTimeDomain,okunoTimedomain}: a Pockels cell and a quarter-wave plate (QWP) sandwiched between two polarizing beam splitters. Another set of these optics forms the optical switch, where the QWP makes the two incident beams interfere when no voltage is applied to the Pockels cell. This is intended to make it possible to sense and stabilize the phase difference between them. The structure of the loop-shaped optical path and the configuration of the homodyne measurement part are the same as those in Ref.~\cite{okunoTimedomain}.

Figure \ref{fig: concept}c shows a typical timing chart of the dynamical components. All the components have the rise/fall time of \SI{\sim10}{\nano\second}, in some cases followed by the transient oscillations of \SI{\sim20}{\nano\second} around the target steady values. The loop-shaped circuit is \SI{18.2}{\metre} long and has the corresponding round-trip time ($\tau$) of \SI{60.8}{\nano\second}, which determines the interval between the neighboring optical pulses to be \SI{60.8}{\nano\second}. Consequently, each component is controlled every \SI{60.8}{\nano\second}.

The phase conditions of the entire optical system are stabilized with auxiliary classical beams. Injecting the classical beams enables us to obtain the error signals and stabilize the system by using the feedback controls. During the main measurement, the auxiliary beams are chopped and the whole system is kept in the stabilized state. The phase of the ancillary squeezed vacuum with respect to that of the input cat state is changed by inverting the corresponding feedback polarity.

\section{Data analysis}
\noindent We experimentally determine the three parameters $\gamma_1$, $\gamma_2$, and $t_0$ in Eq. \eqref{eq: modefunction} representing the mode function of the cat state. As described in Sec. \ref{subsec: single-step gate}, we acquire a time series of homodyne data containing the generated cat state before performing the squeezing gates. From the time series, the three parameters are optimized to maximize the quadrature variance, based on the core idea in Ref.~\cite{morinExperimentally}. The determined values for $\gamma_1$ and $\gamma_2$ are $2\pi\times\SI{29.8}{\mega\hertz}$ and $2\pi\times\SI{95.6}{\mega\hertz}$, respectively.

Each temporal mode in the time-domain-multiplexed processing is defined as $f(t-(i-1)\tau)$ ($i=1,2,\dots$), where $f(t)$ denotes the function of Eq. \eqref{eq: modefunction} with the above-determined parameters. The mode index $i$ corresponds to that shown in Fig. \ref{fig: concept}(a). The raw voltage data from the oscilloscope is converted to the quadrature amplitude by normalizing it by that of the vacuum field obtained immediately before the main measurement.

To fully characterize the gate outcome, we repeat the gate and the measurement for the measurement basis from \ang{90} to \ang{-75} at \ang{15} intervals. \num{3000} independent data are acquired for each basis. The Wigner function is reconstructed from the $\num{3000}\times\num{12}$ quadratures by the maximum-likelihood method~\cite{lvovskyIterative}.

\section{Theoretical models} \label{app: theoretical models}
\noindent To derive the theoretical output Wigner functions shown in Figs. \ref{fig: result1} and \ref{fig: result2}, we employ two different models: the realistic model with finite squeezing level of the ancillae and optical loss, and the ideal model with infinite squeezing level and no optical loss.

The realistic model emulates the experiment as follows: (1) the input cat state is coupled to the finitely squeezed vacuum at the VBS; (2) one output of the VBS is immediately measured at the HD with unity efficiency; (3) the other output is subjected to the propagation efficiency of \num{0.96} in the loop and followed by the measurement at the HD with unity efficiency. For a multi-step gate, the whole process is repeated with the input state replaced with the previous output. For the emulation of squeezing gates with positive (negative) squeezing parameters, we use the experimentally obtained Wigner function in Fig. \ref{fig: result1}(a) (Fig. \ref{fig: result1}(d)) for the initial cat state, and the ancillary squeezed vacua are numerically reproduced by applying optical loss of \SI{22}{\percent} (\SI{27}{\percent}) to the pure squeezed vacua with a squeezing level of \SI{-6.8}{\decibel} (\SI{-7.0}{\decibel}). The amount of loss and the pure squeezing level were estimated from the measurement data simultaneously obtained with that of the input cat. Note that, in this model, the first-step squeezing introduces another \SI{4}{\percent} loss on the ancilla since its implementation obliges the state to pass through the loop to interfere with the input. The effect of non-unity efficiency of the homodyne measurement is assumed to be reflected on the measured input cat and ancilla. The model does not include other experimental imperfections such as the non-ideal phase difference between the cat and the ancilla and the spatial mode mismatch at the VBS between the incoming and the circulated beams.

The ideal model has infinitely squeezed ancilla and no optical loss, deriving the output Wigner function by simply rescaling the experimentally obtained input Wigner function by $e^{-r}$ and $e^{r}$ along the $x$ and $p$ axis, respectively, with the target squeezing parameter $r$.

We also use these models to calculate the output quadrature variances from the experimentally measured input ones.

\add{\section{Current limitations and future prospects}}
\add{\subsection{Optical loss} \label{app: optical loss}}
\add{
    The intrinsic optical losses in the current system are as follows. Based on a preliminary measurement of the squeezing spectrum, we estimate the optical loss between OPA-1 and the HD in Fig. \ref{fig: concept}(d) to be \SI{\sim30}{\percent}, comprising \SI{\sim12}{\percent} internal loss in the OPA module~\cite{kashiwazakiFabrication}, \SI{\sim8}{\percent} in-fiber propagation loss, \SI{\sim3}{\percent} free-space propagation loss, and other unidentified losses. Similarly, the loss between OPA-2 and the HD is estimated to be \SI{\sim20}{\percent}, including \SI{\sim12}{\percent} (or less, due to the absence of fiber coupling at the output) internal loss in the module and \SI{\sim7}{\percent} free-space propagation loss. The roundtrip loss in the loop is measured at \SI{4}{\percent}, and the readout loss at the HD is \SI{\sim3}{\percent}.
}

\add{
    During the squeezing gate process, the input cat state suffers propagation loss in the loop, which degrades the gate quality, as indicated by the reduction in the negativity of the Wigner function and the fidelity $F_{\mathrm{ideal}}$. In addition, the losses on the ancillary squeezed vacuum constrain its squeezing level to \SI{\sim-4}{\decibel}, introducing excess noise with the corresponding variance into one of the quadratures of the gate output state (see Eq. (2) in Ref.~\cite{filipMeasurementinduced}). This effect also contributes to the degradation of the gate performance.
}

\add{
    Let us then discuss the strategy to mitigate these losses in future experiments. For the light source, the OPA module's loss could be reduced by improving the fabrication technique of waveguide nonlinear crystals or eliminating extra optical components within the module~\cite{kashiwazakiFabrication,kashiwazakiOver8dB}. Regarding the measurement system, using an OPA followed by a broadband HD could enable low-loss and high-bandwidth measurements~\cite{inoueMulticore}. Such broadband measurements would reduce the optical pulse length. This reduction would eventually shorten the loop and consequently reduce the optical loss in it, since the minimal loop length is limited by the optical pulse length as well as the electric switching time of the VBS.
}

\newadd{
    In Figs. \ref{fig: result2}(k) and \ref{fig: result2}(l), we present two theoretical lines illustrating potential improvements in gate quality achieved by reducing the loop loss and enhancing the squeezing level of the ancillary squeezed vacuum. The dashdot orange line reflects the effect of approximately halving both the loop loss and the ancilla noise from their current amounts described above. The dotted green line depicts an outcome under a quarter of the current loop loss and \SI[number-unit-product=\text{-}]{15}{\decibel} squeezed ancilla noise, which is the highest recorded squeezing level to date~\cite{vahlbruchDetection}. In this latter scenario, if the loss during non-Gaussian state preparation is also reduced to \SI{13}{\percent}, as achieved in a similar state generation experiment yielding the minimum Wigner negativity~\cite{kawasakiGeneration}, a squeezing gate with a squeezing parameter of $r=0.1,\ 0.2,\ 0.3,\ 0.4$ could be iterated up to \num{18}, \num{10}, \num{7}, and \num{5} times, respectively, while preserving the negative region of the output Wigner function. This level of scalability is realistically anticipated in our system.
}

\add{\subsection{Clock frequency}}
\add{
    We here discuss the practical temporal limitations of our architecture. The current system has the optical pulse interval of \SI{60.8}{\nano\second} or the clock frequency of \SI{16}{\mega\hertz}. This interval is limited by the sum of the optical pulse length of \SI{\sim20}{\nano\second} and the dynamical components' response time of \SI{\sim30}{\nano\second}. The latter is currently limited by the VBS's response, which includes the rise/fall time of \SI{\sim10}{\nano\second} and the subsequent oscillation lasting \SI{\sim20}{\nano\second} (see Fig. \ref{fig: concept}(c)), and demands a buffer time of a few nanoseconds to compensate for the electric jitter.
}

\add{
    The pulse length is currently restricted by the bandwidth of the HD, \SI{\sim200}{\mega\hertz}. This length can be reduced by adopting highly broadband measurements leveraging an OPA, enabling a bandwidth exceeding \SI{10}{\giga\hertz}~\cite{inoueMulticore}. The response rate of the VBS can also be improved to several tens of \si{\giga\hertz} by using a waveguide modulator in place of the current bulk Pockels cell. Therefore, with these enhancements, a clock frequency of \SI{10}{\giga\hertz} or higher is feasible with our architecture.
}

\add{\subsection{Scalability with probabilistically generated non-Gaussian states}}
\add{
    Even if a platform can process non-Gaussian states in a scalable manner, the probabilistic nature of non-Gaussian state generation often limits the overall efficiency of computing protocols involving these states, thereby hindering scalability. However, a methodology has been reported that enables scalable computing with probabilistically generated non-Gaussian states, as long as the generation rate exceeds a certain threshold~\cite{bourassaBlueprint}. The generation rate can be enhanced by broadening the state's bandwidth, employing a more sophisticated generation scheme~\cite{tomodaBoosting}, or multiplexing the non-Gaussian light source~\cite{collinsIntegrated,kanedaTimemultiplexed}. Therefore, our platform has the potential to eventually achieve true scalability.
}


\begin{thebibliography}{10}
\expandafter\ifx\csname url\endcsname\relax
    \def\url#1{\texttt{#1}}\fi
\expandafter\ifx\csname urlprefix\endcsname\relax\def\urlprefix{URL }\fi
\providecommand{\bibinfo}[2]{#2}
\providecommand{\eprint}[2][]{\url{#2}}

\bibitem{bruzewiczTrappedion}
\bibinfo{author}{Bruzewicz, C.~D.}, \bibinfo{author}{Chiaverini, J.},
    \bibinfo{author}{McConnell, R.} \& \bibinfo{author}{Sage, J.~M.}
\newblock \bibinfo{title}{Trapped-ion quantum computing: {{Progress}} and
    challenges}.
\newblock \emph{\bibinfo{journal}{Applied Physics Reviews}}
    \textbf{\bibinfo{volume}{6}}, \bibinfo{pages}{021314} (\bibinfo{year}{2019}).

\bibitem{googlequantumaiSuppressing}
\bibinfo{author}{{Google Quantum AI}} \emph{et~al.}
\newblock \bibinfo{title}{Suppressing quantum errors by scaling a surface code
    logical qubit}.
\newblock \emph{\bibinfo{journal}{Nature}} \textbf{\bibinfo{volume}{614}},
    \bibinfo{pages}{676--681} (\bibinfo{year}{2023}).

\bibitem{winterspergerNeutral}
\bibinfo{author}{Wintersperger, K.} \emph{et~al.}
\newblock \bibinfo{title}{Neutral atom quantum computing hardware: Performance
    and end-user perspective}.
\newblock \emph{\bibinfo{journal}{EPJ Quantum Technology}}
    \textbf{\bibinfo{volume}{10}}, \bibinfo{pages}{1--26} (\bibinfo{year}{2023}).

\bibitem{takedaLargescale}
\bibinfo{author}{Takeda, S.} \& \bibinfo{author}{Furusawa, A.}
\newblock \bibinfo{title}{Toward large-scale fault-tolerant universal photonic
    quantum computing}.
\newblock \emph{\bibinfo{journal}{APL Photonics}} \textbf{\bibinfo{volume}{4}},
    \bibinfo{pages}{060902} (\bibinfo{year}{2019}).

\bibitem{yoshikawaDemonstration}
\bibinfo{author}{Yoshikawa, J.-i.} \emph{et~al.}
\newblock \bibinfo{title}{Demonstration of deterministic and high fidelity
    squeezing of quantum information}.
\newblock \emph{\bibinfo{journal}{Physical Review A}}
    \textbf{\bibinfo{volume}{76}}, \bibinfo{pages}{060301}
    (\bibinfo{year}{2007}).

\bibitem{ukaiDemonstration}
\bibinfo{author}{Ukai, R.}, \bibinfo{author}{Yokoyama, S.},
    \bibinfo{author}{Yoshikawa, J.-i.}, \bibinfo{author}{{van Loock}, P.} \&
    \bibinfo{author}{Furusawa, A.}
\newblock \bibinfo{title}{Demonstration of a {{Controlled-Phase Gate}} for
    {{Continuous-Variable One-Way Quantum Computation}}}.
\newblock \emph{\bibinfo{journal}{Physical Review Letters}}
    \textbf{\bibinfo{volume}{107}}, \bibinfo{pages}{250501}
    (\bibinfo{year}{2011}).

\bibitem{suGate}
\bibinfo{author}{Su, X.} \emph{et~al.}
\newblock \bibinfo{title}{Gate sequence for continuous variable one-way quantum
    computation}.
\newblock \emph{\bibinfo{journal}{Nature Communications}}
    \textbf{\bibinfo{volume}{4}}, \bibinfo{pages}{2828} (\bibinfo{year}{2013}).

\bibitem{asavanantTimeDomainMultiplexed}
\bibinfo{author}{Asavanant, W.} \emph{et~al.}
\newblock \bibinfo{title}{Time-{{Domain-Multiplexed Measurement-Based Quantum
    Operations}} with 25-{{MHz Clock Frequency}}}.
\newblock \emph{\bibinfo{journal}{Physical Review Applied}}
    \textbf{\bibinfo{volume}{16}}, \bibinfo{pages}{034005}
    (\bibinfo{year}{2021}).

\bibitem{larsenDeterministic}
\bibinfo{author}{Larsen, M.~V.}, \bibinfo{author}{Guo, X.},
    \bibinfo{author}{Breum, C.~R.}, \bibinfo{author}{{Neergaard-Nielsen}, J.~S.}
    \& \bibinfo{author}{Andersen, U.~L.}
\newblock \bibinfo{title}{Deterministic multi-mode gates on a scalable photonic
    quantum computing platform}.
\newblock \emph{\bibinfo{journal}{Nature Physics}}
    \textbf{\bibinfo{volume}{17}}, \bibinfo{pages}{1018--1023}
    (\bibinfo{year}{2021}).

\bibitem{enomotoProgrammable}
\bibinfo{author}{Enomoto, Y.}, \bibinfo{author}{Yonezu, K.},
    \bibinfo{author}{Mitsuhashi, Y.}, \bibinfo{author}{Takase, K.} \&
    \bibinfo{author}{Takeda, S.}
\newblock \bibinfo{title}{Programmable and sequential {{Gaussian}} gates in a
    loop-based single-mode photonic quantum processor}.
\newblock \emph{\bibinfo{journal}{Science Advances}}
    \textbf{\bibinfo{volume}{7}}, \bibinfo{pages}{eabj6624}
    (\bibinfo{year}{2021}).

\bibitem{yonezuTimeDomain}
\bibinfo{author}{Yonezu, K.}, \bibinfo{author}{Enomoto, Y.},
    \bibinfo{author}{Yoshida, T.} \& \bibinfo{author}{Takeda, S.}
\newblock \bibinfo{title}{Time-{{Domain Universal Linear-Optical Operations}}
    for {{Universal Quantum Information Processing}}}.
\newblock \emph{\bibinfo{journal}{Physical Review Letters}}
    \textbf{\bibinfo{volume}{131}}, \bibinfo{pages}{040601}
    (\bibinfo{year}{2023}).

\bibitem{bartlettEfficient}
\bibinfo{author}{Bartlett, S.~D.}, \bibinfo{author}{Sanders, B.~C.},
    \bibinfo{author}{Braunstein, S.~L.} \& \bibinfo{author}{Nemoto, K.}
\newblock \bibinfo{title}{Efficient {{Classical Simulation}} of {{Continuous
    Variable Quantum Information Processes}}}.
\newblock \emph{\bibinfo{journal}{Physical Review Letters}}
    \textbf{\bibinfo{volume}{88}}, \bibinfo{pages}{097904}
    (\bibinfo{year}{2002}).

\bibitem{miyataImplementation}
\bibinfo{author}{Miyata, K.} \emph{et~al.}
\newblock \bibinfo{title}{Implementation of a quantum cubic gate by an adaptive
    non-{{Gaussian}} measurement}.
\newblock \emph{\bibinfo{journal}{Physical Review A}}
    \textbf{\bibinfo{volume}{93}}, \bibinfo{pages}{022301}
    (\bibinfo{year}{2016}).

\bibitem{baragiolaAllGaussian}
\bibinfo{author}{Baragiola, B.~Q.}, \bibinfo{author}{Pantaleoni, G.},
    \bibinfo{author}{Alexander, R.~N.}, \bibinfo{author}{Karanjai, A.} \&
    \bibinfo{author}{Menicucci, N.~C.}
\newblock \bibinfo{title}{All-{{Gaussian Universality}} and {{Fault Tolerance}}
    with the {{Gottesman-Kitaev-Preskill Code}}}.
\newblock \emph{\bibinfo{journal}{Physical Review Letters}}
    \textbf{\bibinfo{volume}{123}}, \bibinfo{pages}{200502}
    (\bibinfo{year}{2019}).

\bibitem{yamasakiCostreduced}
\bibinfo{author}{Yamasaki, H.}, \bibinfo{author}{Matsuura, T.} \&
    \bibinfo{author}{Koashi, M.}
\newblock \bibinfo{title}{Cost-reduced all-{{Gaussian}} universality with the
    {{Gottesman-Kitaev-Preskill}} code: {{Resource-theoretic}} approach to cost
    analysis}.
\newblock \emph{\bibinfo{journal}{Physical Review Research}}
    \textbf{\bibinfo{volume}{2}}, \bibinfo{pages}{023270} (\bibinfo{year}{2020}).

\bibitem{miwaExploring}
\bibinfo{author}{Miwa, Y.} \emph{et~al.}
\newblock \bibinfo{title}{Exploring a {{New Regime}} for {{Processing Optical
    Qubits}}: {{Squeezing}} and {{Unsqueezing Single Photons}}}.
\newblock \emph{\bibinfo{journal}{Physical Review Letters}}
    \textbf{\bibinfo{volume}{113}}, \bibinfo{pages}{013601}
    (\bibinfo{year}{2014}).

\bibitem{wangExperimental}
\bibinfo{author}{Wang, M.} \emph{et~al.}
\newblock \bibinfo{title}{Experimental {{Preparation}} and {{Manipulation}} of
    {{Squeezed Cat States}} via an {{All-Optical In-Line Squeezer}}}.
\newblock \emph{\bibinfo{journal}{Laser \& Photonics Reviews}}
    \textbf{\bibinfo{volume}{16}}, \bibinfo{pages}{2200336}
    (\bibinfo{year}{2022}).

\bibitem{okunoTimedomain}
\bibinfo{author}{Okuno, D.} \emph{et~al.}
\newblock \bibinfo{title}{Time-domain programmable beam-splitter operations for
    an optical phase-sensitive non-{{Gaussian}} state}.
\newblock \emph{\bibinfo{journal}{Physical Review A}}
    \textbf{\bibinfo{volume}{110}}, \bibinfo{pages}{023706}
    (\bibinfo{year}{2024}).

\bibitem{filipMeasurementinduced}
\bibinfo{author}{Filip, R.}, \bibinfo{author}{Marek, P.} \&
    \bibinfo{author}{Andersen, U.~L.}
\newblock \bibinfo{title}{Measurement-induced continuous-variable quantum
    interactions}.
\newblock \emph{\bibinfo{journal}{Physical Review A}}
    \textbf{\bibinfo{volume}{71}}, \bibinfo{pages}{042308}
    (\bibinfo{year}{2005}).

\bibitem{neergaard-nielsenGeneration}
\bibinfo{author}{{Neergaard-Nielsen}, J.~S.}, \bibinfo{author}{Nielsen, B.~M.},
    \bibinfo{author}{Hettich, C.}, \bibinfo{author}{M{\o}lmer, K.} \&
    \bibinfo{author}{Polzik, E.~S.}
\newblock \bibinfo{title}{Generation of a {{Superposition}} of {{Odd Photon
    Number States}} for {{Quantum Information Networks}}}.
\newblock \emph{\bibinfo{journal}{Physical Review Letters}}
    \textbf{\bibinfo{volume}{97}}, \bibinfo{pages}{083604}
    (\bibinfo{year}{2006}).

\bibitem{wakuiPhoton}
\bibinfo{author}{Wakui, K.}, \bibinfo{author}{Takahashi, H.},
    \bibinfo{author}{Furusawa, A.} \& \bibinfo{author}{Sasaki, M.}
\newblock \bibinfo{title}{Photon subtracted squeezed states generated with
    periodically poled {{KTiOPO}}\_4}.
\newblock \emph{\bibinfo{journal}{Optics Express}}
    \textbf{\bibinfo{volume}{15}}, \bibinfo{pages}{3568} (\bibinfo{year}{2007}).

\bibitem{herriottOffAxis}
\bibinfo{author}{Herriott, D.}, \bibinfo{author}{Kogelnik, H.} \&
    \bibinfo{author}{Kompfner, R.}
\newblock \bibinfo{title}{Off-{{Axis Paths}} in {{Spherical Mirror
    Interferometers}}}.
\newblock \emph{\bibinfo{journal}{Applied Optics}}
    \textbf{\bibinfo{volume}{3}}, \bibinfo{pages}{523--526}
    (\bibinfo{year}{1964}).

\bibitem{lvovskyIterative}
\bibinfo{author}{Lvovsky, A.~I.}
\newblock \bibinfo{title}{Iterative maximum-likelihood reconstruction in
    quantum homodyne tomography}.
\newblock \emph{\bibinfo{journal}{Journal of Optics B: Quantum and
    Semiclassical Optics}} \textbf{\bibinfo{volume}{6}},
    \bibinfo{pages}{S556--S559} (\bibinfo{year}{2004}).

\bibitem{jozsaFidelity}
\bibinfo{author}{Jozsa, R.}
\newblock \bibinfo{title}{Fidelity for {{Mixed Quantum States}}}.
\newblock \emph{\bibinfo{journal}{Journal of Modern Optics}}
    \textbf{\bibinfo{volume}{41}}, \bibinfo{pages}{2315--2323}
    (\bibinfo{year}{1994}).

\bibitem{ourSupplementalMaterial}
\bibinfo{note}{See Supplemental Material for additional experimental
    results, which includes Ref.~\cite{maurodarianoParameter}.}

\bibitem{maurodarianoParameter}
\bibinfo{author}{Mauro~D'Ariano, G.}, \bibinfo{author}{Paris, M. G.~A.} \&
    \bibinfo{author}{Sacchi, M.~F.}
\newblock \bibinfo{title}{Parameter estimation in quantum optics}.
\newblock \emph{\bibinfo{journal}{Physical Review A}}
    \textbf{\bibinfo{volume}{62}}, \bibinfo{pages}{023815}
    (\bibinfo{year}{2000}).

\bibitem{yukawaEmulating}
\bibinfo{author}{Yukawa, M.} \emph{et~al.}
\newblock \bibinfo{title}{Emulating quantum cubic nonlinearity}.
\newblock \emph{\bibinfo{journal}{Physical Review A}}
    \textbf{\bibinfo{volume}{88}}, \bibinfo{pages}{053816}
    (\bibinfo{year}{2013}).

\bibitem{konnoLogical}
\bibinfo{author}{Konno, S.} \emph{et~al.}
\newblock \bibinfo{title}{Logical states for fault-tolerant quantum computation
    with propagating light}.
\newblock \emph{\bibinfo{journal}{Science}} \textbf{\bibinfo{volume}{383}},
    \bibinfo{pages}{289--293} (\bibinfo{year}{2024}).

\bibitem{sakaguchiNonlinear}
\bibinfo{author}{Sakaguchi, A.} \emph{et~al.}
\newblock \bibinfo{title}{Nonlinear feedforward enabling quantum computation}.
\newblock \emph{\bibinfo{journal}{Nature Communications}}
    \textbf{\bibinfo{volume}{14}}, \bibinfo{pages}{3817} (\bibinfo{year}{2023}).

\bibitem{lloydQuantum}
\bibinfo{author}{Lloyd, S.} \& \bibinfo{author}{Braunstein, S.~L.}
\newblock \bibinfo{title}{Quantum {{Computation}} over {{Continuous
    Variables}}}.
\newblock \emph{\bibinfo{journal}{Physical Review Letters}}
    \textbf{\bibinfo{volume}{82}}, \bibinfo{pages}{1784--1787}
    (\bibinfo{year}{1999}).

\bibitem{takedaUniversal}
\bibinfo{author}{Takeda, S.} \& \bibinfo{author}{Furusawa, A.}
\newblock \bibinfo{title}{Universal {{Quantum Computing}} with
    {{Measurement-Induced Continuous-Variable Gate Sequence}} in a {{Loop-Based
    Architecture}}}.
\newblock \emph{\bibinfo{journal}{Physical Review Letters}}
    \textbf{\bibinfo{volume}{119}}, \bibinfo{pages}{120504}
    (\bibinfo{year}{2017}).

\bibitem{kashiwazakiFabrication}
\bibinfo{author}{Kashiwazaki, T.} \emph{et~al.}
\newblock \bibinfo{title}{Fabrication of low-loss quasi-single-mode {{PPLN}}
    waveguide and its application to a modularized broadband high-level
    squeezer}.
\newblock \emph{\bibinfo{journal}{Applied Physics Letters}}
    \textbf{\bibinfo{volume}{119}}, \bibinfo{pages}{251104}
    (\bibinfo{year}{2021}).

\bibitem{kashiwazakiOver8dB}
\bibinfo{author}{Kashiwazaki, T.} \emph{et~al.}
\newblock \bibinfo{title}{Over-8-{{dB}} squeezed light generation by a
    broadband waveguide optical parametric amplifier toward fault-tolerant
    ultra-fast quantum computers}.
\newblock \emph{\bibinfo{journal}{Applied Physics Letters}}
    \textbf{\bibinfo{volume}{122}}, \bibinfo{pages}{234003}
    (\bibinfo{year}{2023}).

\bibitem{takedaOndemand}
\bibinfo{author}{Takeda, S.}, \bibinfo{author}{Takase, K.} \&
    \bibinfo{author}{Furusawa, A.}
\newblock \bibinfo{title}{On-demand photonic entanglement synthesizer}.
\newblock \emph{\bibinfo{journal}{Science Advances}}
    \textbf{\bibinfo{volume}{5}}, \bibinfo{pages}{eaaw4530}
    (\bibinfo{year}{2019}).

\bibitem{morinExperimentally}
\bibinfo{author}{Morin, O.}, \bibinfo{author}{Fabre, C.} \&
    \bibinfo{author}{Laurat, J.}
\newblock \bibinfo{title}{Experimentally {{Accessing}} the {{Optimal Temporal
    Mode}} of {{Traveling Quantum Light States}}}.
\newblock \emph{\bibinfo{journal}{Physical Review Letters}}
    \textbf{\bibinfo{volume}{111}}, \bibinfo{pages}{213602}
    (\bibinfo{year}{2013}).

\bibitem{inoueMulticore}
\bibinfo{author}{Inoue, A.} \emph{et~al.}
\newblock \bibinfo{title}{Toward a multi-core ultra-fast optical quantum
    processor: 43-{{GHz}} bandwidth real-time amplitude measurement of 5-{{dB}}
    squeezed light using modularized optical parametric amplifier with {{5G}}
    technology}.
\newblock \emph{\bibinfo{journal}{Applied Physics Letters}}
    \textbf{\bibinfo{volume}{122}}, \bibinfo{pages}{104001}
    (\bibinfo{year}{2023}).

\bibitem{vahlbruchDetection}
\bibinfo{author}{Vahlbruch, H.}, \bibinfo{author}{Mehmet, M.},
    \bibinfo{author}{Danzmann, K.} \& \bibinfo{author}{Schnabel, R.}
\newblock \bibinfo{title}{Detection of 15 {{dB Squeezed States}} of {{Light}}
    and their {{Application}} for the {{Absolute Calibration}} of {{Photoelectric
    Quantum Efficiency}}}.
\newblock \emph{\bibinfo{journal}{Physical Review Letters}}
    \textbf{\bibinfo{volume}{117}}, \bibinfo{pages}{110801}
    (\bibinfo{year}{2016}).

\bibitem{kawasakiGeneration}
\bibinfo{author}{Kawasaki, A.} \emph{et~al.}
\newblock \bibinfo{title}{Generation of highly pure single-photon state at
    telecommunication wavelength}.
\newblock \emph{\bibinfo{journal}{Optics Express}}
    \textbf{\bibinfo{volume}{30}}, \bibinfo{pages}{24831--24840}
    (\bibinfo{year}{2022}).

\bibitem{bourassaBlueprint}
\bibinfo{author}{Bourassa, J.~E.} \emph{et~al.}
\newblock \bibinfo{title}{Blueprint for a {{Scalable Photonic Fault-Tolerant
    Quantum Computer}}}.
\newblock \emph{\bibinfo{journal}{Quantum}} \textbf{\bibinfo{volume}{5}},
    \bibinfo{pages}{392} (\bibinfo{year}{2021}).

\bibitem{tomodaBoosting}
\bibinfo{author}{Tomoda, H.} \emph{et~al.}
\newblock \bibinfo{title}{Boosting the generation rate of squeezed
    single-photon states by generalized photon subtraction}.
\newblock \emph{\bibinfo{journal}{Physical Review A}}
    \textbf{\bibinfo{volume}{110}}, \bibinfo{pages}{033717}
    (\bibinfo{year}{2024}).

\bibitem{collinsIntegrated}
\bibinfo{author}{Collins, M.~J.} \emph{et~al.}
\newblock \bibinfo{title}{Integrated spatial multiplexing of heralded
    single-photon sources}.
\newblock \emph{\bibinfo{journal}{Nature Communications}}
    \textbf{\bibinfo{volume}{4}}, \bibinfo{pages}{2582} (\bibinfo{year}{2013}).

\bibitem{kanedaTimemultiplexed}
\bibinfo{author}{Kaneda, F.} \emph{et~al.}
\newblock \bibinfo{title}{Time-multiplexed heralded single-photon source}.
\newblock \emph{\bibinfo{journal}{Optica}} \textbf{\bibinfo{volume}{2}},
    \bibinfo{pages}{1010--1013} (\bibinfo{year}{2015}).

\end{thebibliography}

\end{document}


\newcounter{num} 

\title{
  {\mdseries Supplemental Material\\for}\\
  \newadd{Sequential and Programmable Squeezing Gates}\\\newadd{for Optical Non-Gaussian Input States}
}

\author{Takato Yoshida}
\author{Daichi Okuno}
\author{Takahiro Kashiwazaki}
\author{Takeshi Umeki}
\author{\\Shigehito Miki}
\author{Fumihiro China}
\author{Masahiro Yabuno}
\author{Hirotaka Terai}
\author{Shuntaro Takeda}

\maketitle
\section{\texorpdfstring{Squeezing gates on the vacuum field}{I.. SQUEEZING GATES ON THE VACUUM FIELD}}
\noindent For the reliable evaluation of the squeezing gates, we also apply the same gates as in the main text to the vacuum field. Just by chopping the pump beam of the OPA-1 in Fig. 1(d) in the main text, the input state is replaced with the vacuum while other working conditions are left unchanged.

\begin{figure}[H]
    \centering
    \includegraphics[scale=1]{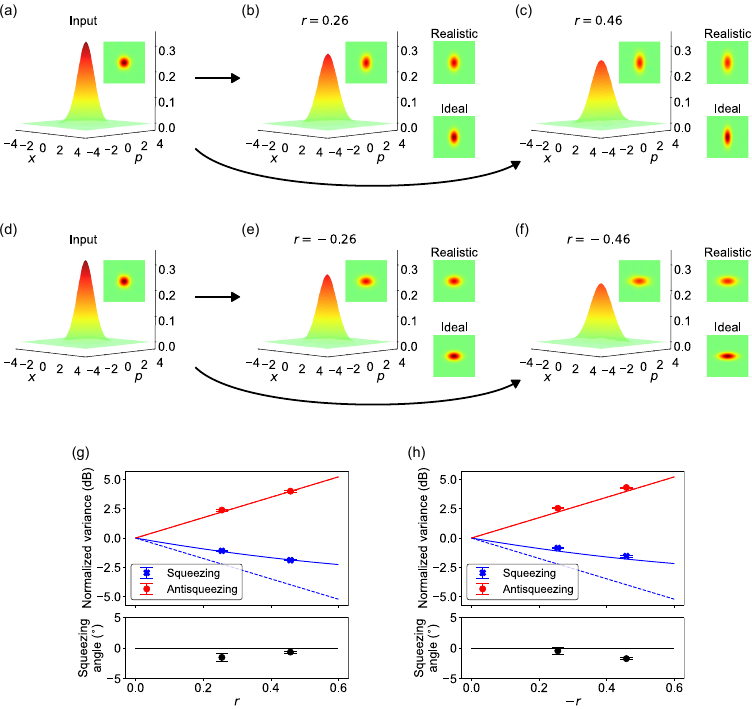}
    \caption{\ \ State characterizations \newerase{of}\newadd{for} the single-step squeezing gates with the input being the vacuum field. (a),(d) \newadd{The }Wigner function of the input vacuum field for the squeezing\newadd{,} with positive and negative squeezing parameters, respectively (\newadd{in principle, }they are identical\newerase{ in principle}). The convention $\hbar=1$ is adopted. (b),(c),(e),(f) \newadd{The }Wigner function of the experimentally obtained output state and the theoretically derived \newerase{ones}\newadd{states}. The horizontal and the vertical directions of the inset square panel correspond to $x$ and $p$, respectively. \newerase{See}\newadd{For} the \newerase{main text for the }description of the theoretical predictions\newadd{, see the main text}. (g),(h) \newerase{Parameters}\newadd{The parameters} about the output ellipse as a function of a target squeezing parameter. The solid and the dashed lines are derived from the realistic and the ideal models, respectively. \newerase{See}\newadd{For} the \newerase{text for the }definition of each parameter\newadd{, see the text}. The center values and the error ranges of the parameters are estimated as averages and standard errors when splitting the whole \num{36000} data into five independent subsets, respectively. The same is the case for Fig. \hyperref[fig: result_vac2_top]{\ref{fig: result_vac2}}.}
    \label{fig: result_vac1}
\end{figure}

\begin{figure}[H]
    \centering
    \includegraphics[scale=1]{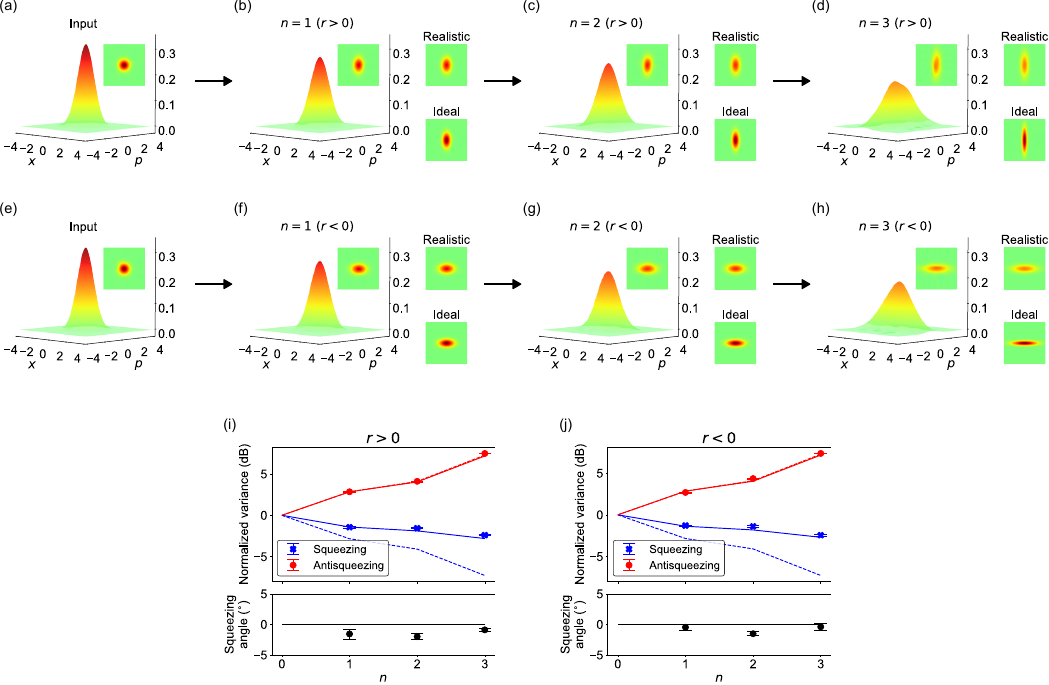}
    \caption{\ \ State characterizations \newerase{of}\newadd{for} the \newerase{multi-step}\newadd{multistep} squeezing gates with the input being the vacuum field. (a),(e) \newadd{The }Wigner function of the input vacuum field for the squeezing\newadd{,} with positive and negative squeezing parameters, respectively. These are the same figures as in Fig. \hyperref[fig: result_vac1_top]{\ref{fig: result_vac1}}. The convention $\hbar=1$ is adopted. (b),(c),(d),(f),(g),(h) \newadd{The }Wigner function of the experimentally obtained output state and the theoretically derived \newerase{ones}\newadd{states}. $n$ denotes the number of steps. (i),(j) \newerase{Parameters}\newadd{The parameters} about the output ellipse as a function of the number of steps. All items are shown in the same manner as in Fig. \hyperref[fig: result_vac1_top]{\ref{fig: result_vac1}}\newadd{,} except that (i) and (j) are shown as functions of $n$.}
    \label{fig: result_vac2}
\end{figure}

\newerase{Figures }\neweraseforref{\ref{fig: result_vac1}, \ref{fig: result_vac2}, along with Tables \ref{fig: result_vac1}, and \ref{fig: result_vac2}}\newerase{, summarize the results of the single-step and the multi-step gates.}\newadd{The results of the single-step and the multistep gates are summarized in Figs. \hyperref[fig: result_vac1_top]{\ref{fig: result_vac1}}, \hyperref[fig: result_vac2_top]{\ref{fig: result_vac2}}, Tables \hyperref[tab: condition_vac1_top]{\ref{tab: condition_vac1}}, and \hyperref[tab: condition_vac2_top]{\ref{tab: condition_vac2}}.} In this experiment, the theoretical outputs are derived with the input state being the ideal vacuum for simplicity. These figures and tables are shown in the same manner as the counterparts in the main text except for Figs. \hyperref[fig: result_vac1_top]{\ref{fig: result_vac1}(g)}, \hyperref[fig: result_vac1_top]{\ref{fig: result_vac1}(h)}, \hyperref[fig: result_vac2_top]{\ref{fig: result_vac2}(i)}, and \hyperref[fig: result_vac2_top]{\ref{fig: result_vac2}(j)} representing the quantitative parameters. Only for these four figures, we evaluate output states assuming that the Wigner function is Gaussian~\cite{maurodarianoParameter}. This assumption is valid since a Gaussian gate does not transform the input Gaussian state into a non-Gaussian state. With this assumption, the contour of the output Wigner function is an ellipse that we call an output ellipse. The normalized quadrature variance along its minor (major) axis and its tilt angle represent the squeezing (antisqueezing) level and the squeezing angle, respectively, since the ellipse of the input vacuum is a circle. The normalization factor for the quadrature variance is $1/2$ as we adopt the convention $\hbar=1$. As evident in Figs. \hyperref[fig: result_vac1_top]{\ref{fig: result_vac1}}, \hyperref[fig: result_vac2_top]{\ref{fig: result_vac2}}, along with Tables \hyperref[tab: condition_vac1_top]{\ref{tab: condition_vac1}}, and \hyperref[tab: condition_vac2_top]{\ref{tab: condition_vac2}}, all the experimental outcomes show little \newerase{deviations}\newadd{deviation} from the theoretical predictions, supporting the validity of the gates of our platform.

\newpage

\phantomsection
\label{tab: condition_vac1_top}
\begin{table}[H]
    \caption{\ \ \newerase{Working}\newadd{The working} conditions and fidelities for the single-step squeezing gates with the input being the vacuum field. All items are listed in the same manner as in Table \setcounter{num}{1}\Roman{num} in the main text.}
    \label{tab: condition_vac1}
    \begin{center}
      \begin{tabularx}{\textwidth}{LCCCCCCCc}
        \hline\hline \\
        Gate & & \multicolumn{4}{c}{Working condition} & \multicolumn{2}{c}{\erase{Result}\add{Experimental fidelities}} & \add{Theoretical fidelity} \\
        \multicolumn{2}{l}{\rule{0.2\textwidth}{0.03em}} & \multicolumn{4}{c}{\rule{0.4\textwidth}{0.03em}} & \multicolumn{2}{c}{\rule{0.2\textwidth}{0.03em}} & \multicolumn{1}{r}{\rule{0.14\textwidth}{0.03em}} \\
        $r$ & Squeezed & Ancilla & $R_1$ & $\phi_1$ & $g_1$ & $F_{\mathrm{real}}$ & $F_{\mathrm{ideal}}$ & \add{$F_{\mathrm{ideal}}^{\mathrm{(theory)}}$} \\
            & quadrature &       &       & \newadd{(degrees)} & &                 &                      & \\
        \hline \\
        \num{0.26} & $\hat{x}$ & $\hat{x}$-sq. & \num{0.40} & \newadd{\num{90}} & \num{-0.82} & \num{0.9919(3)} & \num{0.922(8)} & \add{\num{0.937}} \\
        \num{0.46} & $\hat{x}$ & $\hat{x}$-sq. & \num{0.60} & \newadd{\num{90}} & \num{-1.22} & \num{0.9916(9)} & \num{0.868(6)} & \add{\num{0.872}} \\
        \num{-0.26} & $\hat{p}$ & $\hat{p}$-sq. & \num{0.40} & \newadd{\num{0}} & \num{-0.82} & \num{0.9905(9)} & \num{0.899(3)} & \add{\num{0.934}} \\
        \num{-0.46} & $\hat{p}$ & $\hat{p}$-sq. & \num{0.60} & \newadd{\num{0}} & \num{-1.22} & \num{0.9894(3)} & \num{0.831(4)} & \add{\num{0.868}} \\
        \hline\hline
      \end{tabularx}
    \end{center}
\end{table}

\phantomsection
\label{fig: result_vac1_top}

\newpage

\phantomsection
\label{tab: condition_vac2_top}
\begin{table}[H]
    \caption{\ \ \newerase{Working}\newadd{The working} conditions and fidelities for the \newerase{multi-step}\newadd{multistep} squeezing gates with the input being the vacuum field. All items are listed in the same manner as in Table \setcounter{num}{2}\Roman{num} in the main text.}
    \label{tab: condition_vac2}
    \begin{center}
      {\scriptsize
      \begin{tabularx}{\textwidth}{lCCcccCCc}
        \hline\hline \\
        Gate & & \multicolumn{4}{c}{Working condition} & \multicolumn{2}{c}{\erase{Result}\add{Experimental fidelities}} & \add{Theoretical fidelity} \\
        \multicolumn{2}{l}{\rule{0.24\textwidth}{0.03em}} & \multicolumn{4}{c}{\rule{0.42\textwidth}{0.03em}} & \multicolumn{2}{c}{\rule{0.16\textwidth}{0.03em}} & \multicolumn{1}{r}{\rule{0.13\textwidth}{0.03em}} \\
        $(r_1,r_2,\dots)$ & Squeezed & \newadd{Ancillas} & $(R_1,R_2,\dots)$ & $(\phi_1,\phi_2,\dots)$ & $(g_1,g_2,\dots)$ & $F_{\mathrm{real}}$ & $F_{\mathrm{ideal}}$ & \add{$F_{\mathrm{ideal}}^{\mathrm{(theory)}}$} \\
        & quadrature &       &       & \newadd{(degrees)} & &                 &                      & \\
        \hline \\
        $(\num{0.33})$ & $\hat{x}$ & $\hat{x}$-sq. & $(\num{0.48})$ & \newadd{$(\num{90})$} & $(\num{-0.96})$ & \num{0.991(2)} & \num{0.914(5)} & \add{\num{0.915}} \\
        $(\num{0.33},\num{0.14})$ & $\hat{x}$ & $\hat{x}$-sq. & $(\num{0.48},\num{0.75})$ & \newadd{$(\num{90},\num{90})$} & $(\num{-0.96},\num{0.58})$ & \num{0.9905(9)} & \num{0.848(2)} & \add{\num{0.870}} \\
        $(\num{0.33},\num{0.14},\num{0.37})$ & $\hat{x}$ & $\hat{x}$-sq. & $(\num{0.48},\num{0.75},\num{0.48})$ & \newadd{$(\num{90},\num{90},\num{90})$} & $(\num{-0.96},\num{0.58},\num{1.04})$ & \num{0.977(2)} & \num{0.68(1)} & \add{\num{0.732}} \\
        $(\num{-0.33})$ & $\hat{p}$ & $\hat{p}$-sq. & $(\num{0.48})$ & \newadd{$(\num{0})$} & $(\num{-0.96})$ & \num{0.989(1)} & \num{0.909(7)} & \add{\num{0.912}} \\
        $(\num{-0.33},\num{-0.14})$ & $\hat{p}$ & $\hat{p}$-sq. & $(\num{0.48},\num{0.75})$ & \newadd{$(\num{0},\num{0})$} & $(\num{-0.96},\num{0.58})$ & \num{0.987(1)} & \num{0.818(4)} & \add{\num{0.866}} \\
        $(\num{-0.33},\num{-0.14},\num{-0.37})$ & $\hat{p}$ & $\hat{p}$-sq. & $(\num{0.48},\num{0.75},\num{0.48})$ & \newadd{$(\num{0},\num{0},\num{0})$} & $(\num{-0.96},\num{0.58},\num{1.04})$ & \num{0.981(2)} & \num{0.692(8)} & \add{\num{0.725}} \\
        \hline\hline
      \end{tabularx}
      }
    \end{center}
\end{table}

\phantomsection
\label{fig: result_vac2_top}
